\documentclass[12pt]{article}
\usepackage[left=25mm,top=22mm,right=25mm,bottom=22mm]{geometry}
\usepackage[utf8]{inputenc}

\usepackage[sort, numbers]{natbib}
\newcommand{\msun}{{$M_\odot$}}

\usepackage{caption}
\usepackage{upgreek} 
\usepackage{textgreek}
\usepackage{longtable}
\usepackage{multicol}
\usepackage{subfig}
\usepackage{float}
\usepackage{amsmath}
\usepackage{multirow}
\usepackage{pdfpages}
\usepackage{emptypage}
\usepackage[utf8]{inputenc}
\usepackage{wrapfig}
\usepackage{amssymb}
\usepackage{lineno}
\usepackage{graphicx}

\usepackage{rotating}
\usepackage{hyperref}
\usepackage{xcolor}
\usepackage{etoolbox}
\newcommand{\mb}[1] {{\color{black}\bf#1}}

\newcommand{\SRG}{{\em SRG}}

\makeatletter
\renewcommand{\section}{\@startsection%
{section}{1}{0mm}{-\baselineskip}%
{0.5\baselineskip}{\normalfont\large\bfseries}}
\makeatother

\usepackage{hyperref}
\usepackage{xcolor}
\hypersetup{
  colorlinks   = true, 
  urlcolor     = black, 
  linkcolor    = black, 
  citecolor   = blue 
}
\usepackage[all]{hypcap} 

\input{insbox}

\begin{document}

\bigskip
\bigskip
\bigskip
\bigskip

\begin{center} 
\Large{\bf A snapshot of the oldest AGN feedback phases}
\end{center}

\bigskip
\bigskip

\noindent {\bf M. Brienza$^{1,2}$, 
T. W. Shimwell$^{3,4}$,
F. de Gasperin$^{2,5}$,
I. Bikmaev$^{6,7}$,
A. Bonafede$^{1,2,5}$,
A. Botteon$^{4}$,
M. Br\"uggen$^{5}$,
G. Brunetti$^{2}$,
R. Burenin$^{8}$,
A. Capetti$^{9}$,
E. Churazov$^{8,10}$,
M. J. Hardcastle$^{11}$,
I. Khabibullin$^{8,10}$,
N. Lyskova$^{8}$,
H. J. A. R\"ottgering$^{4}$,
R. Sunyaev$^{8,10}$,
R. J. van Weeren$^{4}$,
F. Gastaldello$^{12}$,
S. Mandal$^{4}$,
S. J. D. Purser$^{13}$,
A. Simionescu$^{14,4,15}$,
C. Tasse$^{16,17,18}$
}

\bigskip

\noindent\emph{$^1$ Dipartimento di Fisica e Astronomia, Università di Bologna, Via P. Gobetti 93/2, I-40129, Bologna, Italy\\
$^2$ INAF - Istituto di Radio Astronomia, Via P. Gobetti 101, I-40129 Bologna, Italy\\
$^3$ ASTRON, Netherlands Institute for Radio Astronomy, Oude Hoogeveensedijk 4, 7991 PD, Dwingeloo, The Netherlands\\
$^4$ Leiden Observatory, Leiden University, PO Box 9513, 2300 RA Leiden, The Netherlands\\
$^5$ Hamburger Sternwarte, Universit\"at Hamburg, Gojenbergsweg 112, 21029, Hamburg, Germany\\
$^6$ Department of Astronomy and Satellite Geodesy, Kazan Federal University, Kremlevskaya Str 18, 420008 Kazan, Russia\\
$^7$ Academy of Sciences of Tatarstan, Bauman Str., 20, Kazan, Russia \\  
$^8$ Space Research Institute (IKI), Russian Academy of Sciences, Profsoyuznaya 84/32, 117997 Moscow, Russia\\
$^9$ INAF - Osservatorio Astrofisico di Torino, Strada Osservatorio 20, I-10025 Pino Torinese, Italy\\
$^{10}$ Max Planck Institute for Astrophysics, Karl-Schwarzschild-Str. 1, Garching b. M\"unchen 85741, Germany\\
$^{11}$ Centre for Astrophysics Research, University of Hertfordshire, College Lane, Hatfield AL10 9AB, UK\\
$^{12}$ INAF- Istituto di Astrofisica Spaziale e Fisica Cosmica (IASF) - Milano, Via A. Corti 12, I-20133 Milano, Italy\\
$^{13}$Dublin Institute for Advanced Studies, Astronomy \& Astrophysics Section, 31 Fitzwilliam Place, Dublin 2, D02 XF86, Ireland\\
$^{14}$ SRON Netherlands Institute for Space Research, Sorbonnelaan 2, 3584 CA Utrecht, The Netherlands  \\
$^{15}$Kavli Institute for the Physics and Mathematics of the Universe (WPI), The University of Tokyo, Kashiwa, Chiba 277-8583, Japan\\
$^{16}$ GEPI, Observatoire de Paris, CNRS, Université Paris Diderot, 5 place Jules Janssen, Meudon, France.\\ 
$^{17}$ USN, Observatoire de Paris, CNRS, PSL, UO, Nançay, France.\\ 
$^{18}$Centre for Radio Astronomy Techniques and Technologies, Department of Physics and Electronics, Rhodes University, Grahamstown 6140, South Africa}

\newpage

{\bf Active Galactic Nuclei (AGN) inject large amounts of energy into their host galaxies and surrounding environment, shaping their properties and evolution \cite{mcnamara2007, fabian2012}. In particular, AGN jets inflate cosmic-ray lobes, which can rise buoyantly as light `bubbles' in the surrounding medium \cite{gull1973}, displacing and heating the encountered thermal gas and thus halting its spontaneous cooling. These bubbles have been identified in a wide range of systems \cite{werner2019, dunn2005}. However, due to the short synchrotron lifetime of electrons, the most advanced phases of their evolution have remained observationally unconstrained, preventing us to fully understand their coupling with the external medium, and thus AGN feedback. Simple subsonic hydrodynamic models \cite{churazov2000, bruggen2003} predict that the pressure gradients, naturally present around the buoyantly rising bubbles, transform them into toroidal structures, resembling mushroom clouds in a stratified atmosphere. The way and timescales on which these tori will eventually disrupt depend on various factors including magnetic fields and plasma viscosity \cite{reynolds2005, ruszkowski2008}. Here we report observations below 200 MHz, sensitive to the oldest radio-emitting particles, showing the late evolution of multiple generations of cosmic-ray AGN bubbles in a galaxy group with unprecedented level of detail. The bubbles' buoyancy power can efficiently offset the radiative cooling of the intragroup medium. However, the bubbles have still not thoroughly mixed with the thermal gas, after hundreds of million years, likely under the action of magnetic fields.}

\bigskip

The galaxy group is named `Nest200047' and was first identified by \cite{tully2015} (see Supplementary section for more details). At its center lies the massive elliptical galaxy MCG+05-10-007, with spectroscopic redshift $z=0.01795\pm 0.00015$ \cite[ref.][]{huchra2012} (see Supplementary Figure 2). We first discovered the system in the radio band through the {\it Low Frequency Array} (LOFAR, \cite{vanhaarlem2013}) Two-meter Sky Survey (LoTSS, \cite{shimwell2019}) at 144 MHz thanks to the bizarre morphology of its radio emission (see Figure 1 and Supplementary Figure 1). It was then further investigated using follow-up LOFAR observations at 53 MHz and X-ray observations performed with the {\it extended ROentgen Survey with an Imaging Telescope Array} (eROSITA) \cite{predehl2021} on board of the {\it Spectrum-Roentgen-Gamma} (SRG) mission (Sunyaev et al., in prep.). The X-ray emission of the group in the 0.5-2~keV band was clearly detected for the first time by eROSITA (see Figure 2 and Supplementary Figure 3), and appears to be nicely centred at the position of MCG+05-10-007. We estimated that the temperature of the intragroup medium (IGrM) is $kT_X\sim2\pm0.5$~keV and its luminosity is $L_X= 5-10\times 10^{42}\,{\rm erg\,s^{-1}}$. The total mass of the system, lies in the range $M_{500}=3-7\times 10^{13}\, M_{\odot}$, placing Nest200047 among typical groups of galaxies, consistent with the initial classification \cite{tully2015}. 

Our new LOFAR images at 53 MHz and 144 MHz reveal that the galaxy group is permeated by non-thermal plasma with complex morphology extending up to $\sim$200 kpc from the center. In particular, the central galaxy shows AGN activity in the form of two nearly-symmetric radio jets about 25-kpc long. These jets have a radio luminosity equal to $L_{144MHz}=1.7\times10^{23}$ W$\rm Hz^{-1}$ (marked as A in Figure 1, top panels) and a spectral index typical for freshly injected plasma ($\alpha_{53MHz}^{144MHz}(A)\simeq$0.6$\rm \pm 0.18$, $S\propto\nu^{-\alpha}$, where $S$ is the flux density and $\nu$ is the frequency, see Figure 3). As shown in Figure 1, top-right panel, in the innermost regions of the jets we observe two symmetric very compact components, suggesting that the AGN is recurrently inflating new bubbles of radio-emitting plasma. This scenario is also supported by the presence of two extra pairs of low surface brightness lobes located beyond the inner jets (marked as B and C in Figure 1, top panels). These can be interpreted as remnant lobes of past AGN-jet outbursts \cite{schoenmakers2000} or, as a series of bubbles of plasma, which regularly detach from a continuously operating jet. The spectral index gradient observed in these three pairs of lobes
($\alpha_{53MHz}^{144MHz}(B)\simeq$0.8$\rm \pm 0.18$ and $\alpha_{53MHz}^{144MHz}(C)\simeq$1.2$\rm \pm 0.18$) is consistent with electrons being older at larger distances from the central AGN. At a distance of about 200 kpc from the center of the galaxy group, we detect a complex array of filamentary radio-emitting structures connected to the AGN lobes C1 and C2 by bridges of radio emission. These filaments are oriented in various directions and often show sharp bends and double-strands of very narrow emission. Particularly striking are the `box-shaped' filament in the North and the main filament (D2), which extends in the East-West direction for $\sim$350 kpc and has a width of a few kpc, probing the presence of magnetic field coherence on very large scales. A similar, elongated structure is present in the South too, although much fainter and with less defined morphology (marked as D1). Moreover, the LOFAR images at 25 arcsec resolution reveal the presence of diffuse extended emission embedding the filaments, with increasing radio spectral indices (up to $\rm \alpha_{53MHz}^{144MHz}\simeq2.5\pm0.3$), i.e. older emission, towards the source periphery.

\medskip
The presence of multiple generations of AGN lobes and their clear morphological connection with the filamentary structures on larger scales, together with some tentative signs of an X-ray/radio anticorrelation on scales $\sim 100$~kpc, suggest that the recurrent AGN-jet activity is responsible for the creation of the entire observed emission in Nest200047. The edge-brightened rim partly filled with radio-emitting plasma observed in the X-ray image (Figure 2) strongly resembles other AGN feedback-driven objects, such as M84 \cite{finoguenov2001}. In particular, the morphology of the non-thermal plasma in the northern region of the group has a strong resemblance with the `mushroom-shaped' structure observed in M87 \cite{owen2000, churazov2001}, as well as with X-ray cavities in their late phases of evolution as predicted by hydrodynamical simulations (e.g. \cite{yang2019}). Based on this, we interpret the radio-bright structures D2 and D3 as tori of plasma (i.e. vortex rings) expanding in the IGrM as seen approximately edge-on and caught at a much more advanced stage of evolution with respect to M87. The spectral index in the ring D2, reaching values as flat as $\alpha_{53MHz}^{144MHz}\sim 0.75\pm0.2$, might also suggest that a mild compression of the plasma in this region is occurring, for example, as a result of a weak shock launched by a subsequent AGN outburst or of the group dynamics. The effect of this compression would indeed be to increase the synchrotron spectral break frequency by a factor $\sim2.5$ \cite[ref.][]{markevitch2005}, moving it above the currently observed frequency range. Moreover, this might contribute to shaping the morphology of the large-scale emission and to creating the observed filaments, rings and eddies \cite{ensslin2002}. We note that the morphology of D1 and D2 may remind of radio relics in galaxy clusters \cite{rajpurohit2018}. However, the optical (see Sect. 1 in SI) and X-ray data (see Sect. 3 and 4 in Methods) of the system do not support the presence of an ongoing major merger. Furthermore, their spectral index distribution differs from classical relics \cite{rajpurohit2018} and their luminosity falls above the classical correlations found for relics \cite{degasperin2014b}, implying an implausible-high particle acceleration efficiency. All this makes the AGN scenario more likely.

\medskip
Guided by other systems with cavities, we assume that the bubble C2 and D3 were initially formed close to the central AGN and are now rising buoyantly, i.e. subsonically, in the IGrM. The estimated buoyancy ages for these two structures are $t_{buoy, C2} \gtrsim$170 Myr and $t_{buoy, D3}\gtrsim$ 350 Myr. These are consistent with age estimates based on pure particle radiative losses equal to $t_{rad, C2}<350$ Myr, $t_{rad, D3}<400$ Myr or shorter if adiabatic losses are also taken into account. Such high ages have only been found in the Hydra cluster so far \cite{wise2007}, and are about a factor 10 above the mean cavity ages found in galaxy clusters and groups 
\cite{birzan2004}.

Using the buoyancy timescales we estimated the mechanical power deposited into the thermal gas by the bubbles C2 and D3 equal to $\rm P_{bubble,D3}= 1-4\times10^{42}$ erg/s and $\rm P_{bubble,C2}= 3\times10^{41}-1\times10^{42}$ erg/s. The power of the bubble D3 is compatible with the observed X-ray luminosity implying that it might be effective at counterbalancing the IGrM cooling. The lower power of the bubble C2 suggests instead that not all AGN outbursts may have the same energetic impact on the system.

\medskip
A comparison between the thermal pressure of the IGrM as derived from
the X-ray emission (from $\rm p_{th}\sim 2 - 4\times 10^{-12} \ dyne
\ cm^{-2}$ in the core, down to $\rm \sim 10^{-12} \ dyne
  \ cm^{-2}$ near structures D2, D3) and the non-thermal pressure of
the radio-emitting plasma as derived from the minimum energy conditions assuming an electron/positron composition,
($\rm p_{nth} \sim 4\times 10^{-13} \ dyne \ cm^{-2}$ in the core,
  down to $\rm \sim 2 \times 10^{-13} \ dyne \ cm^{-2}$ in D2, D3),
suggests that a non-negligible energetic contribution from
  non-radiating protons in these structures might be present. However,
  such contribution cannot exceed that of radio-emitting electrons by
  a factor larger than $\sim 10$ if the structures are in pressure balance.

\medskip

From hydrodynamic numerical simulations we know that, for a given environment, perfect vortex rings can travel much larger distances with respect to amorphous structures, which tend instead to get shredded after crossing a distance comparable to their size \cite{oneill2009}. We suggest therefore that the observed diffuse emission might be interpreted as the leftover of disrupted bubbles, possibly combined with some level of turbulence, likely injected in the ambient medium by the AGN itself. The radio filaments show major distortions on 100-kpc scales. This might probe shear motions in the IGrM, which eventually cascade into turbulence. Under this assumption we estimate that the energy flux dissipated by the turbulence into IGrM heat is $\rm \sim 4\times 10^{41} \ g \ s^{-3}$. This suggests that a fraction between 1/10 and 1/3 of the bubble $pV$ work (where p is the bubble pressure and V is the volume) is converted into IGrM turbulence, consistent also with turbulence dissipation fractions in merging galaxy clusters \cite{sunyaev2003}. The presence of many thin two-filament structures throughout the system is also consistent with a partially turbulent medium. Indeed, on small scales, magnetohydrodynamic (MHD) turbulence can form a complex network of filamentary structures in field and density fluctuations, which are characterized by dissipative sheet-like structures with magnetic field lines along the long axis confining transverse, tangential shear layers \cite{porter2015}. At these scales, the anisotropic mixing naturally leads to the formation of density filaments along the field lines \cite{xu2019}.

However, the fact that the bubbles and filaments have managed to
maintain their integrity over a travelled distance of 100-200 kpc and
timescales of hundreds of Myr might challenge the presence of a turbulent IGrM. One could expect indeed that the observed narrow filaments (width $h\sim 2-10\,{\rm kpc}$) should have been destroyed by random, turbulent motions if their (1D) velocities $v_h$ at scale $\sim h$ were much larger than $h/t_{buoy}\sim 10-20\,{\rm km\,s^{-1}}$. For comparison, 3D-velocities of $\sim 50\,{\rm km\,s^{-1}}$ on scales 
of $\sim10$~kpc would be needed to balance cooling by turbulent
dissipation \cite{zhuravleva2014}. However, we note that thin
filaments can survive in turbulent environments due to the Reynolds
stress of magnetic fields that make MHD turbulence anisotropic. The
physical scale at which the magnetic tension starts to play a role in
turbulent dynamics is the Alfven scale, which we estimate for this
system to be in the range $\rm l_A=1\sim5$ kpc, consistent with the
width of the filaments. This suggests that magnetic fields might be playing a major role in the stability of the tori/bubbles observed in Nest200047, as also proposed by simulations of magnetised vortex rings \cite{oneill2009, ehlert2018}.

\medskip

Nest200047 clearly offers a rare opportunity to study the evolution of AGN bubbles in a galaxy group over hundreds of Myr, from the ‘inception’ of the youngest pair of lobes all the way to the break-up of the oldest generation of bubbles into the IGrM. Interestingly, despite a long and apparently rather complicated evolution, even the oldest radio plasma is not yet thoroughly mixed with the thermal plasma, neither by diffusion nor by small-scale mixing. However, this lack of mixing by no means reduces the efficiency of the AGN feedback, since the energy exchange between the bubbles and the IGrM can proceed without a thermal coupling of these phases. Thanks to its unprecedented level of detail Nest200047 represents a unique piece of evidence for AGN feedback models and can provide new empirical constraints to magneto-hydrodynamical simulations investigating the coupling between AGN cosmic-ray bubbles and IGrM.

\medskip

\begin{figure*}
\centering
\includegraphics[width=17cm]{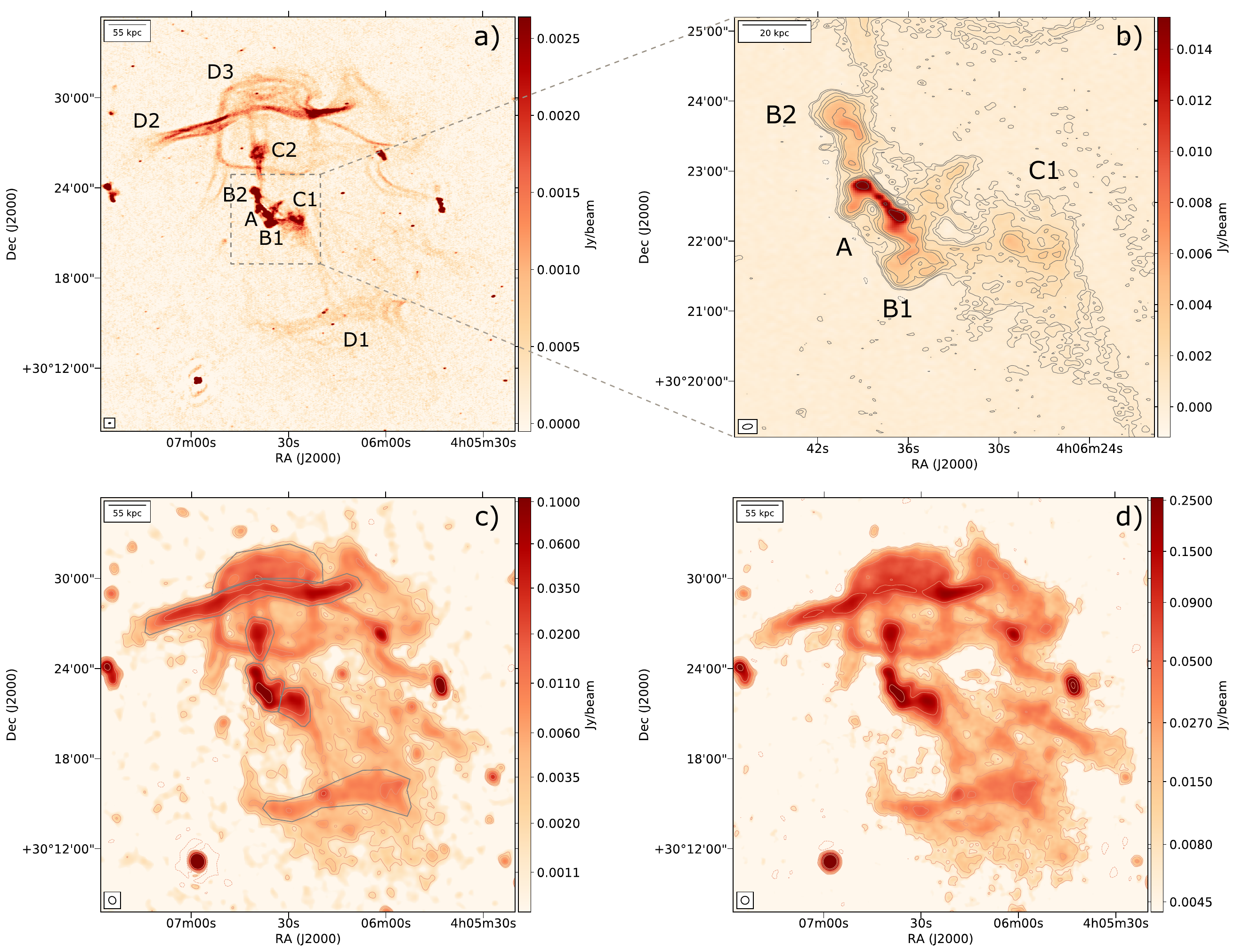}
\caption{\textbf{LOFAR images showing the complex non-thermal radio emission in the galaxy group Nest200047.} \textit{Panel (a):} LOFAR image at 144 MHz with resolution of 4.3 arcsec $\times$ 8.6 arcsec and noise of $\rm \sigma = 0.166 \ mJy \ beam^{-1}$. Letters mark the most relevant morphological features as described in the text. \textit{Panel (b):} Zoom-in on the central AGN at 144 MHz and resolution of 4.3 arcsec $\times$ 8.6 arcsec. Contours are drawn at -3, 3, 5, 10, 15, 20, 30, 100 $\times \sigma$. \textit{Panel (c):} LOFAR image at 144 MHz with resolution of 28.3 arcsec $\times$ 30.5 arcsec and noise of $\rm \sigma = 0.57 \ mJy \ beam^{-1}$. Contours are drawn at -3, 3, 5, 10, 20, 40, 100, 180 $\rm \times \sigma$. \textit{Panel (d):} LOFAR image at 53 MHz with resolution of 32 arcsec and noise of $\rm \sigma = 2.3 \ mJy \ beam^{-1}$. Contours are drawn at -3, 3, 5, 10, 20, 40, 100, 180 $\rm \times \sigma$. The beam size of each map is shown in the bottom left corner of each panel.}
\end{figure*}

\begin{figure*}
\centering
\includegraphics[width=11cm]{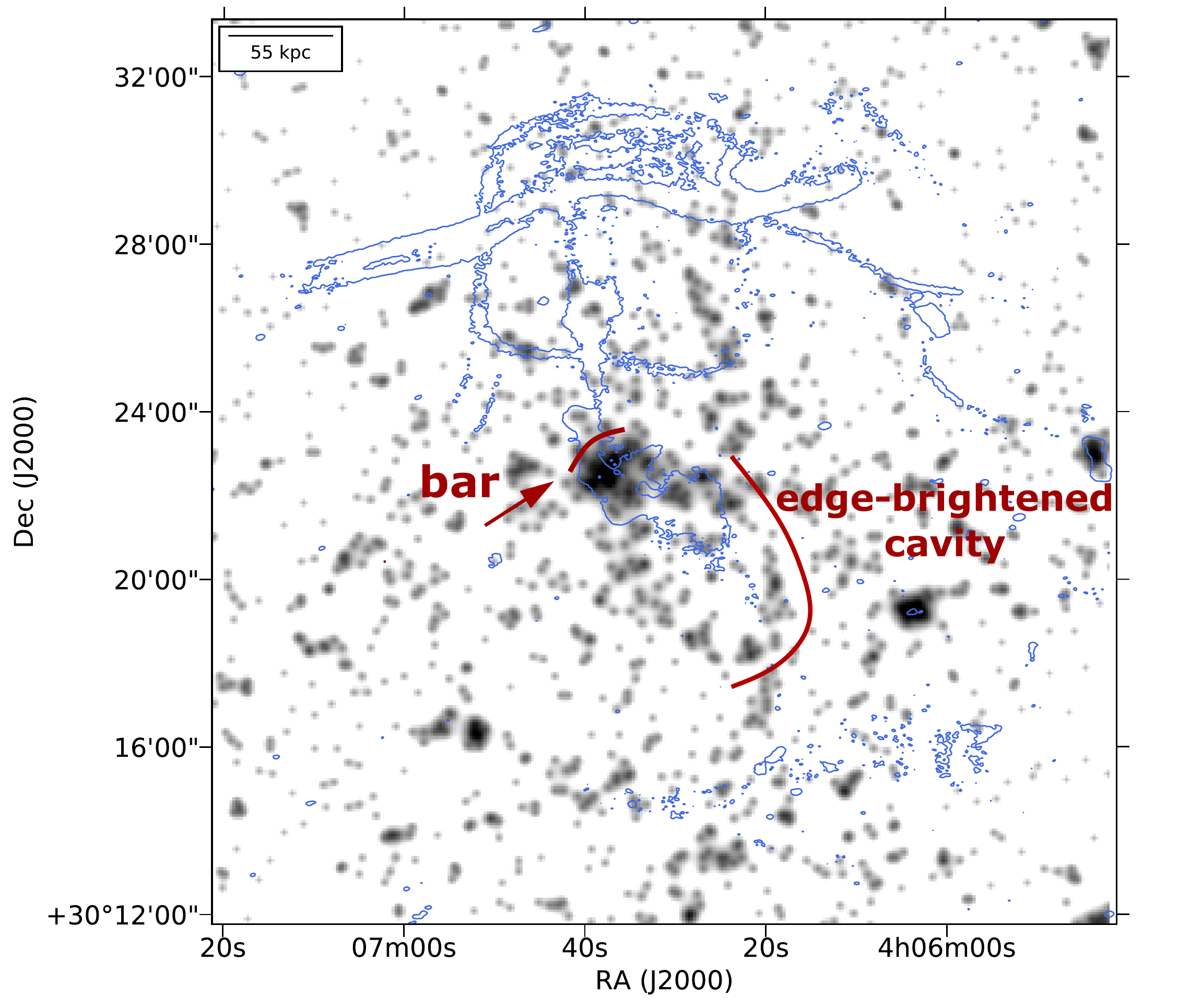}
\caption{\textbf{Lightly smoothed 0.5-2.3 keV eROSITA X-ray image of the galaxy group Nest200047 showing the fine X-ray substructure in the core of the group.} The shape of the smoothing kernel corresponds to the point spread function (PSF) of the telescope (30 arcsec half-power diameter). Overlaid in blue is the 3$\rm\sigma$ radio contour of the LOFAR 144-MHz map at a resolution of 4.3 arcsec $\times$ 8.6 arcsec (as shown in Figure 1). In the core of the galaxy group (central 1 arcmin), an X-ray-bright bar is seen, which is orthogonal to the orientation of the inner radio jets (see red arrow). On slightly larger scales $\sim 5$ arcmin, to the southwest of the nucleus, a limb-brightened X-ray cavity is present (see red line), which is partly filled with the radio lobe C1. The inner 100-kpc region is qualitatively similar to other AGN feedback-driven objects, like, e.g., M84 in the Virgo cluster (e.g., \cite{finoguenov2001})}
\end{figure*}

\begin{figure*}
\centering
\includegraphics[width=11cm]{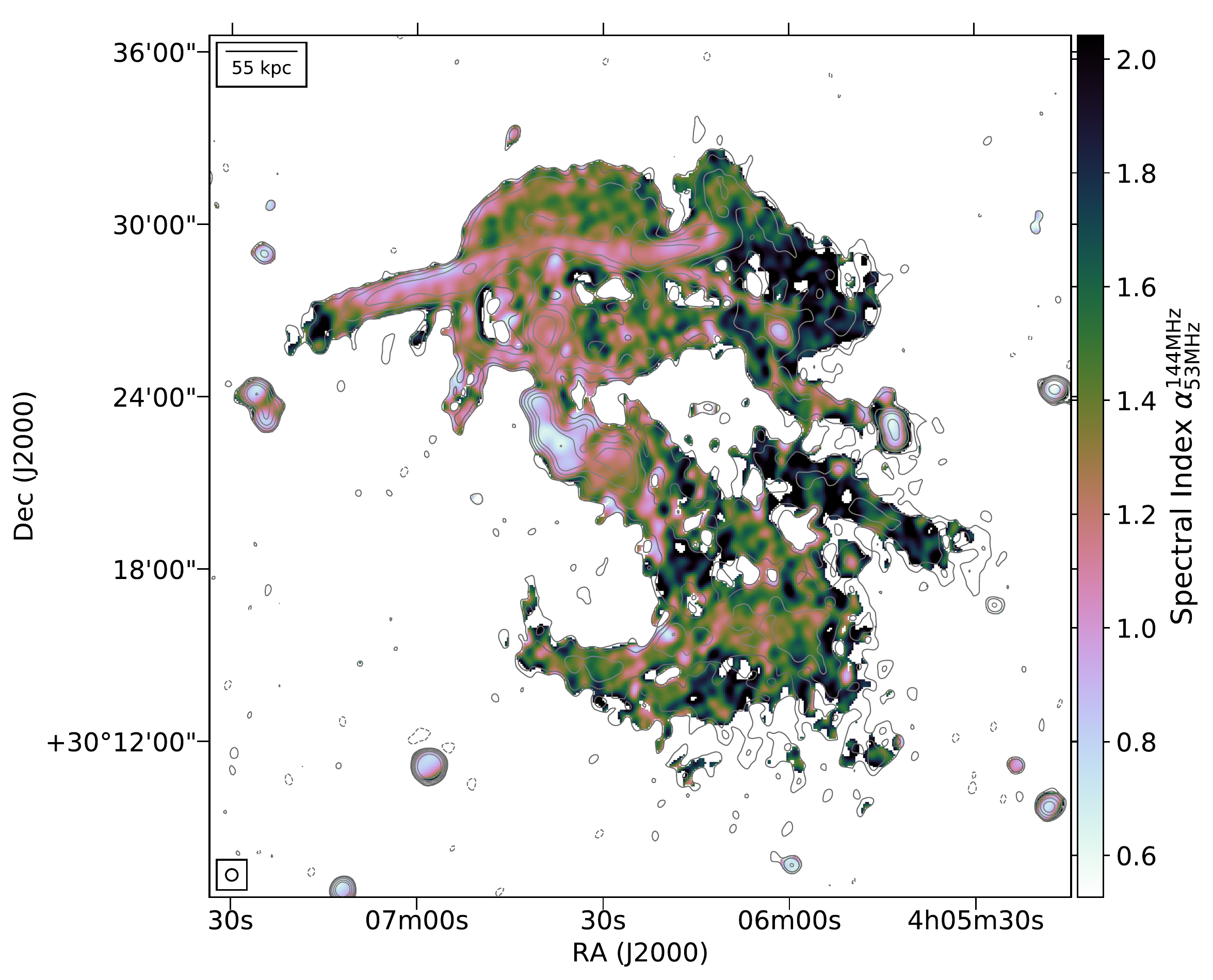}
\caption{\textbf{Spectral index map in the range 53-144 MHz of the galaxy group Nest200047.} The image shows an increasing spectral steepening of the non-thermal plasma towards the system peripheries, implying increasing radiative ages. The map is produced using LOFAR images with uniform weighting scheme, a gaussian uv-taper of 25 arcsec and a restoring beam of 25 arcsec. Only pixels with surface brightness above 3$\rm \sigma$ in both maps have been used. Contour levels represent the emission at 53 MHz and are drawn at -3, 3, 5, 10, 20, 35, 100 $\times \sigma$ levels, with $\sigma$=2.7 mJy $\rm beam^{-1}$. The beam size is shown in the bottom left corner of the image.}
\end{figure*}

\newpage

\noindent {\LARGE \bf Methods}
\bigskip

\noindent{\bf 1. LOFAR observations, data reduction and radio images.} The galaxy group Nest200047 was observed with the LOFAR High-Band Antennas (HBA) with a central frequency of 144 MHz for a total of 16 hours. The observations were performed as part of the LOFAR Two-meter SKy Survey (LoTSS, \cite{shimwell2019}) and pointed in the direction of the two grid pointings P061+29 and P060+31. These have a field of view with full-width at half-maximum (FWHM) of 3.96 deg at 144 MHz \cite{vanhaarlem2013} and their centers lie at 1.44 and 1.32 degrees away from the target position, respectively. The observation setup followed the standard LoTSS strategy, whose main details are summarised in Supplementary Table 1. The data were first flagged for radio frequency interference (RFI) and averaged by a factor 4 in frequency by the observatory before being ingested into the LOFAR long term archive.  The archived data were then corrected for direction-independent effects such as ionospheric Faraday rotation, phase offset between stations correlations (with polarizations XX and YY) and clock offsets (see \cite{degasperin2019}) using the \texttt{PreFactor} pipeline (for more details see https://github.com/lofar-astron/prefactor) as described by \cite{vanweeren2016} and \cite{williams2016}. To correct for ionospheric distortions and errors in the beam model, a direction dependent self-calibration was then performed using DDF-pipeline (for more details see https://github.com/mhardcastle/ddf-pipeline). This pipeline is described by \cite{shimwell2019, tasse2021} and uses \texttt{KillMS} (\cite{tasse2014}, \cite{smirnov2015}) to derive direction dependent calibration solutions, which are then applied during imaging with \texttt{DDFacet} \cite{tasse2018}. Finally, to refine the calibration of the ionosphere and beam in the direction of the target, and to allow for quicker re-imaging, we made use of the post processing procedure described in \cite{vanweeren2020}. Following this, we subtracted from the visibilities all sources outside a box with side=0.5 deg centred on the target and performed a final direction independent self-calibration of the data. The final high resolution image was produced with \texttt{WSClean} (version 2.7, \cite{offringa2014}) using a Briggs weighting scheme with robust=-0.5. 
An inner uv-cut at 40$\rm \lambda$ was also applied to discard very large-scale emission ($\gtrsim$ 1.5 deg), which is unrelated to the target and typically affected by severe RFI and poorly calibrated. The final image has a resolution of 4.3 arcsec $\times$ 8.6 arcsec and an rms of 0.166 mJy $\rm beam^{-1}$ (see Figure 1, top panels). In order to enhance the large-scale diffuse emission of the target, the data were also imaged at lower resolution, using a Briggs weighting scheme with robust=-0.5 and a gaussian uv-taper of 25 arcsesc. The final image has a resolution of 28.5 arcsec $\times$ 32 arcsec and rms of 0.57 mJy$\rm \ beam^{-1}$ (see Figure 1, bottom-left panel).

The target was observed with LOFAR using the Low-Band Antennas (LBA), as part of a Director's Discretionary Time Proposal. We observed for 8 hrs in the frequency range $30 - 78$ MHz with one beam pointed at the calibrator 3C196 and one on the target field. The data were taken at 1s integration time and at frequency resolution of 64 channels per SB (sub band, 0.196 MHz). A summary of the observation setup can be found in Supplementary Table 1. Data affected by RFI were flagged using \texttt{AOflagger} \cite{offringa2012} and subsequently averaged to the time resolution of 2s and the frequency resolution of 0.049 MHz. The calibrator data were reduced following \cite{degasperin2019}. This procedure isolated some systematic effect (polarisation alignment, the bandpass, and the clock drifts), which we transferred to the target data. For the initial direction independent calibration of the target field we followed \cite{degasperin2020}. This procedure removes three systematic effects, averaging them across the entire field of view: the ionospheric delay, the Faraday rotation, and the corrections on the beam variation with time and frequency on top of the LOFAR beam model. The result of this process is a direction-independent calibrated image that reached 3 mJy $\rm beam^{-1}$ rms noise at the resolution of 45 arcsec. Due to direction-dependent errors, largely driven by ionospheric corruption \cite{mevius2016, degasperin2018}, this image shows numerous artefacts, especially around bright sources, which compromise its fidelity. To correct these errors we performed a direction-dependent calibration. The procedure followed the idea outlined in \cite{vanweeren2016}, where a series of local ``direction-dependent" calibrators is located across the field of view. We isolated 7 bright calibrators, including the target source. For each of these sources we did several cycle of phase-only self-calibration. The bright calibrators are finally subtracted from the data accounting for the effect of the ionosphere in their direction. The target direction was processed last and a few cycles of slow amplitude corrections were also performed. As for the higher frequency data we imaged the data with \texttt{WSClean} (version 2.7, \cite{offringa2014}) using a Briggs weighting scheme with robust=-0.5, an inner uv-cut at 40$\rm \lambda$, and a restoring beam of 32 arcsec. The image rms noise is 2.7 mJy$\rm beam^{-1}$ (see Figure 1, bottom-right panel).

The flux scale of both images were checked using as a reference the brightest sources in the field. The flux densities of these sources from all publicly available surveys were used to extrapolate the expected flux densities at the frequencies of interest and these were compared with the measured values. Following this procedure we did not find any systematic offset in the flux scale of any of the two images.

\medskip

\noindent{\bf 2. Radio flux densities and spectral indices.} To measure the radio flux density of the source and to investigate its spectral behaviour we have re-imaged both the 53-MHz and the 144-Hz datasets using a common inner uv-cut at 40$\rm \lambda$, uniform weighting scheme and a gaussian uv-taper of 25 arcsec. The total flux density of the source (measured using the 3$\sigma$ contours as a reference) and the flux densities of the major features within the source (measured using the regions drawn in Figure 1, bottom-left panel) at both frequencies are listed in Supplementary Table 2, together with their respective luminosities and global spectral indices. The total errors on the flux densities were computed by combining in quadrature the flux scale errors (10\% at 53 MHz, \cite{degasperin2021} and 15\% at 144 MHz \cite{shimwell2019}) and the image noise multiplied by the flux density integration area. Flux densities of the main compact sources present in the field (see Supplementary Figure 4) are also reported in Supplementary Table 3. To produce the spectral index map shown in Figure 3, the spectrum of each pixel was fitted with a power-law and only pixels above 3$\sigma$ were included. The errors on the global spectral indices and on the spectral index map (see Supplementary Figure 5) were obtained using the following formula:

\begin{equation}
\rm \alpha_{err} = \frac{1}{ln\frac{53}{144}}\sqrt{\left(\frac{\Delta S_{53}}{S_{53}}\right)^2+\left(\frac{\Delta S_{144}}{S_{144}}\right)^2}
\end{equation}

where $\rm {S_{53}}$ and $\rm {S_{144}}$ are the flux density (surface brightness) values at the respective frequencies and $\rm \Delta S_{53}$ and $\rm \Delta S_{144}$ their corresponding errors. The adopted cosmology is $\rm{\Lambda}$CDM cosmology, with $\rm{\Omega_{m}}$ = 0.3, $\rm{\Omega_\Lambda}$ = 0.7 and H$_{0}$ = 70 km\,s$^{-1}$Mpc$^{-1}$.

\medskip

\noindent{\bf 3. SRG/eROSITA observations, data reduction and images.} The region where the galaxy group Nest200047 is located was routinely scanned by the \SRG observatory (Sunyaev et al. in prep.), observatory, featuring the eROSITA telescope \cite{predehl2021}, in March and September 2020, and February 2021 for a total of $\sim 645$ seconds in the course of the two half-year all-sky surveys. The eROSITA 0.5-2.3 keV images are shown in Figure 2 and Supplementary Figure 3.

In these images, we can clearly see that the diffuse X-ray emission is approximately spherical and centered at the position of MCG+05-10-007, suggesting that the system is in an overall dynamically relaxed state. The X-ray emission in merging systems is indeed expected to show an elongated and/or double-peaked morphology along the merger axis (e.g. \cite{rajpurohit2018}) and is centered on the brightest central galaxy only in rare cases with peculiar projection effects \cite{rossetti2016}.

We note that the brightest central patch of the X-ray emission (better seen in Figure 2) is clearly extended in the direction perpendicular to the orientation of the inner radio jets. Similar structures formed by the cool gas have also been found in the cores of giant elliptical galaxies affected by AGN feedback (see, e.g., \cite{finoguenov2001} for X-ray image of M84). 

The proximity of Nest200047 to the Galactic plane (Galactic latitude $b\approx -16^\circ$) has to be considered when interpreting these data. Indeed, the distribution of neutral gas and dust  \cite{hi4pi2016,planck2014, schlegel1998} suggests (i) variable low-energy photoelectric absorption across the field (see Supplementary Figure 3) and (ii) variable contribution of the Galactic diffuse emission. Both factors can affect the appearance of the diffuse emission, especially on scales $\gg10$ arcmin. On scales less than 10-20 arcmin, the impact of these effects on the 0.5-2.3 keV data is rather modest, but not negligible. This limits the accuracy of the total flux estimates by a factor $\sim 2$.

\medskip

\noindent{\bf 4. X-ray surface brightness profile, luminosity and  mass.} Using the eROSITA map described above we derived the radial profile of the X-ray emission as shown in Supplementary Figure 6. From the plot we can see that the diffuse X-ray emission can be traced up to $\sim$20-30 arcmin from the group center and that an excess in the central regions is present with respect to a beta-model. While deeper observations are required to quantify a possible contribution of the central AGN to this excess, this trend points to the presence of a cool core.

For the spectral analysis we have selected a $13$ arcmin circle around the central AGN. Given the complexity of the foreground and low-energy absorption discussed in Sect. 3, we have experimented with different regions to estimate and subtract the contribution of the foreground spectra. We have concluded that a $13-29$ arcmin annulus is a reasonable choice for the `background' region. The spectra were fitted with the AtomDB/\texttt{APEC} thermal plasma emission model (see http://atomdb.org; \cite{foster2012}). From the X-ray spectra and the optical observations (see Section 1 in SI) the effective hydrogen column density was found to be $N_H\sim 5 \times 10^{21}\,{\rm cm^{-2}}$. The best-fitting value of the gas temperature is $kT_X \sim 2$~keV. As expected, this value depends on the choice of the background region and lies in the range 1.5-2.5 keV. The 0.5-2~keV X-ray luminosity of the group within a 30-arcmin circle (after correction for Milky Way absorption) is $L_X \sim (5-10) \times 10^{42}\,{\rm erg\,s^{-1}}$.

Using the $\beta$-model fit to the surface brightness profile (core-radius $r_c\approx 7$ arcmin, corresponding to a physical size of $\sim$140 kpc, and $\beta=0.64$; see Supplementary Figure 6), one can estimate the gas mass of the group.
For a gas metallicity varying between 0.3 and 1 times the solar value and for an absorption column density of $\sim 5 \times 10^{21}\,{\rm cm^{-2}}$, the resulting $M_{gas}$ varies from $ 2\times 10^{12}\, M_{\odot}$ to $5 \times 10^{12}\, M_{\odot}$. Assuming the gas mass fraction within $R_{500}$ to be $f_{gas} \simeq 0.07$ typical for galaxy groups (e.g. \cite{lovisari2015}), the total mass ($M_{500}$) of the Nest200047 group is then $3\times 10^{13}\, M_{\odot}$ to $7\times 10^{13}\, M_{\odot}$, about a factor of a few lower than the mass derived from the infrared luminosity of the system (see Section 1 in SI).

\medskip

\noindent{\bf 5. Pressure derivation.} To investigate the pressure balance between the non-thermal radio-emitting plasma and its surrounding thermal IGrM we used the following approach. For the non-thermal plasma we made the classical assumption of minimum energy and assumed the magnetic field to be uniformly distributed across the considered volume -- moderate deviations from magnetic field or electron uniformity make little difference to these calculations \cite{hardcastle2013}. For the calculation we used {\sc pysynch} (see https://github.com/mhardcastle/pysynch), which provides a Python interface to the code of \cite{hardcastle1998}. We adopted a Jaffe-Perola \cite{jaffe1973} aged electron spectrum, with a minimum Lorentz factor $\gamma_{\rm min}=1$ and a low-energy injection energy index $q =2\alpha+1=2.2$, which matches the observed low-frequency spectral index in the inner jets, marked as A, as well as being in agreement with observations of Fanaroff-Riley I radio galaxies \cite{carilli1991}. The spectrum was given a radiative age of 250 Myr, assuming a self-consistent loss field strength of 4 $\mu$G and inverse-Compton losses appropriate to the redshift of the source, which gives an approximate fit to the spectral index observed between 53 and 144 MHz for these components (see further spectral ageing analysis below). For true minimum energy we assume an electron-positron plasma, so that there are no non-radiating particles. Volumes are computed assuming a prolate ellipsoidal geometry ($\rm V=4/3\pi a^2b$, where $a$ and $b$ are the ellipsoid semi-axes). We stress, however, that the final pressure values do not significantly depend on the assumed volume (as pressure scales with the volume as $\rm p\propto V^{4/7}$). The final values of volume, magnetic field and pressure are reported in Supplementary Table 4.

We then computed the IGrM thermal pressure using the relation  $p_{th}\approx 1.9 n_e kT$. If we consider an electron density in the core  $\rm n_{e} \sim 0.5 \times 10^{-3} \ cm^{-3}$ and a temperature in the range $\rm kT= 1.5 - 2.5$ keV, as derived from the X-ray data, we get a thermal pressure in the range $\rm p_{th} \simeq (2 -  4)\times 10^{-12} \ dyne \ cm^{-2}$. At the position of the structures D2, D3 (approximately 150 kpc from the group center) we expect that the IGrM pressure should drop by a factor of $\sim$2 to about $\rm p_{th}\sim  10^{-12} \ dyne \ cm^{-2}$.

Overall, the thermal pressure is about an order of magnitude above the non-thermal minimum pressure values shown in Supplementary Table 4, which is similar to the typical ratio of minimum pressure to external thermal pressure seen in FRI radio galaxies \cite{croston2018}. Since the radio structures cannot in reality be underpressured, this implies a non-negligible contribution from non-radiating particles (e.g. protons) which dominate the energetics of the large-scale structures. We note, however, that if the plasma is in pressure balance with the external medium, any possible deviation from equipartition would imply a tighter upper limit on the contribution of non-radiating particles.

\medskip

\noindent{\bf 6. Timescales and powers.} To investigate the age of the non-thermal plasma we estimated the rising time and the radiative age of the two main bubbles C2 and D3.

We computed the rising time using the standard approach t=H/v, where $v$ is the bubble velocity and H is its (projected) height from the group center (e.g. \cite{birzan2004}). An upper limit to the bubble velocity is provided by the sound speed, which can be approached in case the size of the bubble is comparable to the scale-height of the atmosphere. This can be computed as:

\begin{equation}
v_{c_s}=\sqrt{\Gamma \frac{ k  T}{\mu m_p}},  
\end{equation}

where $kT$ is the average IGrM temperature equal to $kT_{X} \sim$ 2 keV as obtained from the X-ray analysis (see Sect. 4), $\mu$=0.62 is the mean molecular weight, $\rm \Gamma$=5/3 is the adiabatic index and $m_p$ is the proton mass. The derived sound speed is equal to $v_{c_s}$=720 km $\rm s^{-1}$, implying minimum rising times equal to $\sim$120 Myr for the bubble C2 and $\sim$240 Myr for the bubble D3, if $H_{C2}\sim90$ kpc and $H_{D3}\sim170$ kpc are assumed.

A more realistic speed value can be obtained based on buoyancy arguments using the following relation:

\begin{equation}
v_{buoy}=\sqrt{\frac{2 g V}{S C}},
\end{equation}

where $g$ is the gravitational acceleration, $V$ the volume of the bubble, $\Phi$ the cross section ($\rm \Phi_{C2}=\pi a^2; \Phi_{D3}=\pi ab$) of the bubble and $C$ the drag coefficient. We assumed $C$=0.75 \cite[ref.][]{churazov2001} and $g=(2 \sigma^2)/H$ \cite{binney1987}, where the velocity dispersion is set to $\sigma$=421 km $\rm s^{-1}$ (see Sect. 1 in SI). The derived buoyancy speeds are $v_{buoy,C2}\sim$670 km $\rm s^{-1}$ and $v_{buoy,D3}\sim$500 km $\rm s^{-1}$, implying rising times of $\rm t_{buoy, C2}\sim$130 Myr for the bubble C2 and $\rm t_{buoy,D3}\sim$350 Myr for the bubble D3. A summary of the bubble properties is presented in Supplementary Table 5. Of course, these age values should be considered as first order estimates. For example, in the case of non-spherical bubbles, the buoyancy speed is expected to be a factor of a few lower than in the idealised case of a spherical shape \cite{zhang2018}. Bubble transport times can further increase in case other processes, such as turbulence and magneto-thermal instabilities play a role in the system. Overall, the presented buoyancy time estimates can safely be considered as lower limits on the bubble age.

Based on the bubble age estimates presented above, the bubble power is calculated as $\rm P_{bubble}=p_{th}V/t_{buoy}$. Assuming pressure values in the range $\rm p_{th}=1-4\times 10^{-12} \ dyne \ cm^{-2}$ (see Section 5) we obtain $\rm P_{bubble,C2}=3\times 10^{41}-1\times 10^{42}$ erg/s and $\rm P_{bubble,D3}=1-4\times 10^{42}$ erg/s.

\smallskip

Using the observed spectral index trend shown in Figure 3, we also derived upper limits to the radiative
  age of the plasma in the different regions of the source. In
  particular, we used the \texttt{BRATS} software \cite{harwood2013},
  which can model radio spectra by integrating numerically the
  radiative age equations (including the radiative losses of the
  plasma through synchrotron emission and inverse-Compton scattering
  with the cosmic microwave background, CMB). We simulated spectra at many age steps using a Jaffe-Perola
  model \cite{jaffe1973} and fixing the magnetic field to the
  conservative value of $B=1/\sqrt(3) \cdot B_{CMB}=1.95$ $\mu$G (which corresponds to the minimum radiative losses allowed for a plasma at a given redshift) and the injection index to the conservative value of $\rm \alpha_{inj}=0.5$ (the lowest allowed by the Fermi theory). By comparing the spectral index in the frequency range 53-144 MHz for each modelled spectrum (each representing a different age) with the empirical spectral index in the same frequency range shown in Figure 3, we inferred that it takes a maximum of 200-300 Myr for the plasma to get a spectral index in the range $\rm \alpha^{53MHz}_{144MHz}=0.75\sim1$ (as observed in D2), 400 Myr to get $\rm \alpha^{53MHz}_{144MHz}\sim1.4$ (as observed in D3) and 350 Myr to get $\rm \alpha^{53MHz}_{144MHz}\sim1.2$ (as observed in C2). As the aforementioned values are computed assuming minimum values of magnetic field and injection index they can be considered as upper limits, which can further reduce if adiabatic losses are taken into account. The radiative values can therefore be considered consistent with the dynamical age presented above.

\medskip
\noindent{\bf 7. Turbulence and energy flux.} If we assume that the curvature observed in the filaments over scales of $\rm L\sim$100 kpc is originated by the shear velocity field, and assuming this corresponds to the turbulence injection scale, we can derive that the turbulence velocity at such scales is $\rm dv_0=L/t_{buoy}\simeq$ 280 km/s (assuming \rm $t_{buoy}$=350 Myr), which corresponds to a Mach number $\rm M=dv_0/v_{c_s}\simeq$0.4. The turbulence velocity at a scale equal to the filaments height (l$\sim$10 kpc) is then:

\begin{equation}
\rm dv(l) = dv_0 \times (l/L)^{1/3} \simeq130 \  km/s 
\end{equation}

When dv(l) reaches the Alfven velocity $\rm v_A$ the tension produced
by the magnetic fields is no longer negligible. In the case of a Kolmogorov cascade the Alfven scale is $\rm l_A = L \times M_A^{-3}$, where $\rm M_A=dv_0/v_A$ is the Alfven Mach number. The Alfven scale can be then written as:

\begin{equation}
\rm l_A = 2 \ kpc \times (t/350\ Myr)^3 \times (L/100 \ kpc)^{-2} \times (\beta_{pl}/100)^{-3/2} .
\end{equation}

where $t$ is the time, $L$ is the length-scale and the plasma beta $\rm
\beta_{pl}=p_{th}/p_{nth} = (6/5)\cdot(v_{c_s}/v_A)^2$. Assuming
t=350 Myr, L=100 kpc and $\beta_{pl}$=50-200 we find that
below a scale of 1$\sim$5 kpc we enter the MHD regime and magnetic
fields can play a role in preventing the filaments from bending.

Finally the energy flux of the turbulence can be computed by using the following expression:

\begin{equation}
\rm f = 1/2 \times \rho \times dv_0^3/L \times V
\end{equation}

From this and assuming the volume of the bubble D3, a velocity of $\rm dv_0=$280 km/s, a density of $\rm \rho= n_e\cdot m_p$ with $n_e=0.5\times10^{-3}$, and a scale of L=100 kpc, we get that the energy flux dissipated by the turbulence into IGrM heat is $\rm f\sim$4 $\rm \times 10^{41} \ g \ s^{-3}$.

\bigskip

\medskip
\noindent{\bf Correspondence.} Correspondence and requests
for materials should be addressed to Marisa Brienza
(m.brienza@ira.inaf.it).

\medskip
\noindent{\bf Acknowledgements.} M. Brienza sincerely thanks F. Santoro, K. Rajpurohit and F. Vazza for their help and very useful discussions. The authors thank the referees for their revision, which helped improving the manuscript. 
M. Brienza and A. Bonafede acknowledge support from the ERC-Stg DRANOEL, no 714245. M. Brienza acknowledges support by the ERC-StG project MAGCOW, no 714196. A. Bonafede acknowledges support from the MIUR grant FARE “SMS”. RJvW acknowledges support from the ERC Starting Grant ClusterWeb 804208. A.Botteon acknowledges support from the VIDI research programme with project number 639.042.729, which is financed by the Netherlands Organisation for Scientific Research (NWO). AS is supported by the Women In Science Excel (WISE) programme of the NWO, and acknowledges the World Premier Research Center Initiative (WPI) and the Kavli IPMU for the continued hospitality. SRON Netherlands Institute for Space Research is supported financially by NWO. M.Br\"uggen acknowledges support from the Deutsche Forschungsgemeinschaft under Germany's Excellence Strategy - EXC 2121 "Quantum Universe" - 390833306. S. J. D. P. would like to acknowledge support from the European Research Council advanced grant H2020–ERC–2016–ADG–74302 under the European Union’s Horizon 2020 Research and Innovation programme. F. Gastaldello and GB acknowledge support from INAF mainstream project ‘Galaxy Clusters Science with LOFAR’ 1.05.01.86.05. IB, RB, and RS thank TUBITAK, IKI, KFU, and AST for partial support in using RTT150 (the Russian-Turkish 1.5-m telescope in Antalya).  The work of IB was funded by the grant 671-2020-0052 to Kazan Federal University.

LOFAR, the Low Frequency Array designed and constructed by ASTRON, has facilities in several countries, which are owned by various parties (each with their own funding sources), and are collectively operated by the International LOFAR Telescope (ILT) foundation under a joint scientific policy. The ILT resources have benefited from the following recent major funding sources: CNRS-INSU, Observatoire de Paris and Universit\'e d'Orl\'eans, France; BMBF, MIWF-NRW, MPG, Germany; Science Foundation Ireland (SFI), Department of Business, Enterprise and Innovation (DBEI), Ireland; NWO, The Netherlands; the Science and Technology Facilities Council, UK; Ministry of Science and Higher Education, Poland; The Istituto Nazionale di Astrofisica (INAF), Italy.

Part of this work was carried out on the Dutch national e-infrastructure with the support of the SURF Cooperative through grant e-infra 160022 \& 160152. The LOFAR software and dedicated reduction packages on \url{https://github.com/apmechev/GRID_LRT} were deployed on the e-infrastructure by the LOFAR e-infragrop, consisting of J.\ B.\ R.\ Oonk (ASTRON \& Leiden Observatory), A.\ P.\ Mechev (Leiden Observatory) and T. Shimwell (ASTRON) with support from N.\ Danezi
(SURFsara) and C.\ Schrijvers (SURFsara). The J\"ulich LOFAR Long Term Archive and the German LOFAR network are both coordinated and operated by the J\"ulich Supercomputing Centre (JSC), and computing resources on the supercomputer JUWELS at JSC were provided by the Gauss Centre for supercomputing e.V. (grant CHTB00) through the John von Neumann Institute for Computing (NIC).

This research made use of the University of Hertfordshire
high-performance computing facility and the LOFAR-UK computing facility located at the University of Hertfordshire and supported by STFC (ST/P000096/1), and of the Italian LOFAR IT computing infrastructure supported and operated by INAF, and by the Physics Department of Turin University (under an agreement with Consorzio Interuniversitario per la Fisica Spaziale) at the C3S Supercomputing Centre, Italy.

This work is based on observations with eROSITA telescope onboard SRG space observatory. The SRG observatory was built by Roskosmos in the interests of the Russian Academy of Sciences represented by its Space Research Institute (IKI) in the framework of the Russian Federal Space Program, with the participation of the Deutsches Zentrum für Luft- und Raumfahrt (DLR). The eROSITA X-ray telescope was built by a consortium of German Institutes led by MPE, and supported by DLR. The SRG spacecraft was designed, built, launched and is operated by the Lavochkin Association and its subcontractors. The science data are downlinked via the Deep Space Network Antennae in Bear Lakes, Ussurijsk, and Baikonur, funded by Roskosmos. The eROSITA data used in this work were converted to calibrated event lists using the eSASS software system developed by the German eROSITA Consortium and analysed using proprietary data reduction software developed by the Russian eROSITA Consortium.

This research made use of APLpy, an open-source plotting package for Python hosted at \url{http://aplpy.github.com}.

\medskip
\noindent{\bf Author Contributions.} M.Brienza coordinated the research, performed the radio imaging and radio analysis and wrote the manuscript. TWS coordinated the LOFAR HBA data processing and helped coordinating the project. FDG led the LOFAR LBA observing proposal and performed the data reduction. A.Bonafede helped with the analysis of the radio data and with coordination of the project. A.Botteon helped with the LOFAR HBA data processing and with the manuscript revision. M.Br\"uggen and GB helped with interpretation of the source and with the manuscript revision. AC contributed with the system identification and the analysis of the optical properties of the system. MJH helped with the LOFAR HBA data processing, with the interpretation of the source and with manuscript revision. EC, IK, NL analyzed the SRG/eROSITA data and contributed to interpretation of the results and writing the manuscript. IB performed the optical observations and analysis of the source and contributed to the interpretation of the results and writing the manuscript. RB, RS contributed to the interpretation of the results and writing the manuscript. RJvW helped with the LOFAR HBA data processing and with the manuscript revision. FG helped with the interpretation if the results and the manuscript revision. SM, AS helped revising the manuscript. CT helped with the LOFAR HBA data processing. SP led the LOFAR HBA observing proposal. 

\medskip
\noindent{\bf Competing interests.} The authors declare no competing financial interests.

\medskip
\noindent{\bf Data availability.} 
The radio observations are available in the LOFAR Long Term Archive (LTA; https://lta.lofar.eu/). The X-ray datasets are not yet publicly available. Their proprietary scientific exploitation rights were granted by the project funding agencies (Roscosmos and DLR) to two consortia led by MPE (Germany) and IKI (Russia), respectively. The SRG--eROSITA all-sky survey data will be released publicly after a minimum period of 2 years. The exact release date for the data belonging to the consortium led by IKI is yet to be decided. All other data and figures within this paper are available from the corresponding
author upon reasonable request.

\medskip
\noindent{\bf Code availability.} The codes that support the figures within this paper and other
findings of this study are available from the corresponding author upon reasonable request.

\newpage

\centerline {\Large \bf SUPPLEMENTARY INFORMATION}
\bigskip
\bigskip
\bigskip

\noindent {\bf M. Brienza$^{1,2}$, 
T. W. Shimwell$^{3,4}$,
F. de Gasperin$^{2,5}$,
I. Bikmaev$^{6,7}$,
A. Bonafede$^{1,2,5}$,
A. Botteon$^{4}$,
M. Br\"uggen$^{5}$,
G. Brunetti$^{2}$,
R. Burenin$^{8}$,
A. Capetti$^{9}$,
E. Churazov$^{8,10}$,
M. J. Hardcastle$^{11}$,
I. Khabibullin$^{8,10}$,
N. Lyskova$^{8}$,
H. J. A. R\"ottgering$^{4}$,
R. Sunyaev$^{8,10}$,
R. J. van Weeren$^{4}$,
F. Gastaldello$^{12}$,
S. Mandal$^{4}$,
S. J. D. Purser$^{13}$,
A. Simionescu$^{14,4,15}$,
C. Tasse$^{16,17,18}$
}

\bigskip

\noindent\emph{$^1$ Dipartimento di Fisica e Astronomia, Università di Bologna, Via P. Gobetti 93/2, I-40129, Bologna, Italy\\
$^2$ INAF - Istituto di Radio Astronomia, Via P. Gobetti 101, I-40129 Bologna, Italy\\
$^3$ ASTRON, Netherlands Institute for Radio Astronomy, Oude Hoogeveensedijk 4, 7991 PD, Dwingeloo, The Netherlands\\
$^4$ Leiden Observatory, Leiden University, PO Box 9513, 2300 RA Leiden, The Netherlands\\
$^5$ Hamburger Sternwarte, Universit\"at Hamburg, Gojenbergsweg 112, 21029, Hamburg, Germany\\
$^6$ Department of Astronomy and Satellite Geodesy, Kazan Federal University, Kremlevskaya Str 18, 420008 Kazan, Russia\\
$^7$ Academy of Sciences of Tatarstan, Bauman Str., 20, Kazan, Russia \\  
$^8$ Space Research Institute (IKI), Russian Academy of Sciences, Profsoyuznaya 84/32, 117997 Moscow, Russia\\
$^9$ INAF - Osservatorio Astrofisico di Torino, Strada Osservatorio 20, I-10025 Pino Torinese, Italy\\
$^{10}$ Max Planck Institute for Astrophysics, Karl-Schwarzschild-Str. 1, Garching b. M\"unchen 85741, Germany\\
$^{11}$ Centre for Astrophysics Research, University of Hertfordshire, College Lane, Hatfield AL10 9AB, UK\\
$^{12}$ INAF- Istituto di Astrofisica Spaziale e Fisica Cosmica (IASF) - Milano, Via A. Corti 12, I-20133 Milano, Italy\\
$^{13}$Dublin Institute for Advanced Studies, Astronomy \& Astrophysics Section, 31 Fitzwilliam Place, Dublin 2, D02 XF86, Ireland\\
$^{14}$ SRON Netherlands Institute for Space Research, Sorbonnelaan 2, 3584 CA Utrecht, The Netherlands  \\
$^{15}$Kavli Institute for the Physics and Mathematics of the Universe (WPI), The University of Tokyo, Kashiwa, Chiba 277-8583, Japan\\
$^{16}$ GEPI, Observatoire de Paris, CNRS, Université Paris Diderot, 5 place Jules Janssen, Meudon, France.\\ 
$^{17}$ USN, Observatoire de Paris, CNRS, PSL, UO, Nançay, France.\\ 
$^{18}$Centre for Radio Astronomy Techniques and Technologies, Department of Physics and Electronics, Rhodes University, Grahamstown 6140, South Africa}

\newpage

\noindent{\Large{\bf Overview of Nest200047}}\\

\noindent The galaxy group `Nest200047' consists of 17 galaxies, has a velocity dispersion of 421 km~s$^{-1}$ and a total mass equal to $1.46\times 10^{14}$ \msun (adjusted to our cosmology) based on the K$\rm _s$-band luminosities of the galaxy members \cite{tully2015}. At the center of the system lies the massive elliptical galaxy MCG+05-10-007 with k-band magnitude equal to $\rm m_k$=8.87 and mass equal to $\rm log M_*$/\msun=11.56. The second most luminous elliptical galaxy within the radius r500 of the group according to \cite{tully2015} is located at RA 61.9248, DEC 30.2627 and has a k-band magnitude equal to $\rm m_k$=9.9. The observed magnitude gap equal to $\rm \Delta m_k=1.03$ between the two brightest ellipticals suggests that the system has not undergone recent major merger events.

From the large errors on the optical line ratios [NII]/H$\rm \alpha$ [OIII]/H$\rm \beta$ derived by \cite{vandenbosch2015}, we infer that MCG+05-10-007 does not show strong and clear emission lines in its optical spectrum. This was confirmed by dedicated spectroscopic observations with the 1.5-m RTT-150 optical telescope on September 28/29, 2020. 
The lack of signs of the AGN emission is not surprising, given that
Nest200047 is $\sim$5 times more distant than the Virgo cluster. By using the extinction-corrected flux we have determined the luminosity integrated within the slit aperture of 2.4 arcsec $\times$ 4.9 arcsec (corresponding to the 0.9 kpc $\times$ 1.8 kpc at the 75.5 Mpc distance, and 366 parsec/arcsec scale) equal to $\sim1.2\times10^{43}$ erg/s. The physical size of that region is much larger than the central 16 pc $\times$ 16 pc around the black hole in studied in M87  \cite{macchetto1997, anderson2018}. As the result, the observed V-band optical emission in MCG+05-10-007 is dominated by some $\sim$4 billions solar type cool stars, which can completely outshine a low luminosity AGN similar to the M87 nucleus.

All this suggests a classification of this radio galaxy as a Low Excitation Galaxy (LEG). The infrared magnitudes of the galaxy in the Wide-Field Infrared Survey Explorer ({\it WISE}, \cite{wright2010}) are W1=10.116, W2=10.163 and W3=9.655 at 3.4, 4.6 and 12 $\mu$m, respectively. This implies infrared colors equal to W1$-$W2=$-$0.047 and W2-W3=0.508, which locate the galaxy in the region of the {\it WISE} color-color plot occupied by low-excitation radio galaxies \cite{gurkan2014}. Since the Nest200047 is located close to the Galactic plane, the interstellar reddening in this direction is substantial. We derived E(B-V)=0.6 mag by comparing RTT-150 spectrum of MCG+05-10-007 with the template spectrum of a similar type elliptical galaxy with a very small reddening. This E(B-V) value corresponds to N (HI) = $\rm 5.3\times 10^{21} cm^{-2}$, in excellent agreement with N(HI) derived from eROSITA X-ray data.

To date, no dedicated radio observations of Nest 200047 are available in the archives of other telescopes.  Most of the major public radio surveys cover this area, such as the VLA Low-Frequency Sky Survey at 74 MHz, (VLSSr, beam=74 arcsec $\rm \times$ 74 arcsec, $\rm \sigma_{local}=90 \ mJy \ beam^{-1}$; \cite{cohen2007, lane2012}), the Westerbork Northern Sky Survey at 325 MHz (WENSS, beam=74 arcsec $\rm \times$ 107 arcsec, $\rm \sigma_{local}=5 \ mJy \ beam^{-1}$; \cite{rengelink1997}), NVSS at 1400 MHz (beam= 45 arcsec $\times$ 45 arcsec, $\rm \sigma_{local}=0.6 \ mJy \ beam^{-1}$) and the GaLactic and Extragalactic All-sky MWA survey in the frequency range 72-231 MHz (GLEAM,beam=3.5 arcmin $\rm \times$ 2.3 arcmin, at 200 MHz, $\rm \sigma_{local}=50 \ mJy \ beam^{-1}$; \cite{wayth2015}). However, due to their low angular resolution and sensitivity, these images do not resolve the extended radio emission observed in the group or they only recover central radio galaxy. Only in WENSS clear hints of the structures D1, D2, D3 are visible at low signal-to-noise, but the quality of the image does not allow us to perform a detailed analysis of these.

\begin{figure*}[!htp]
\centering
{\includegraphics[width=16cm]{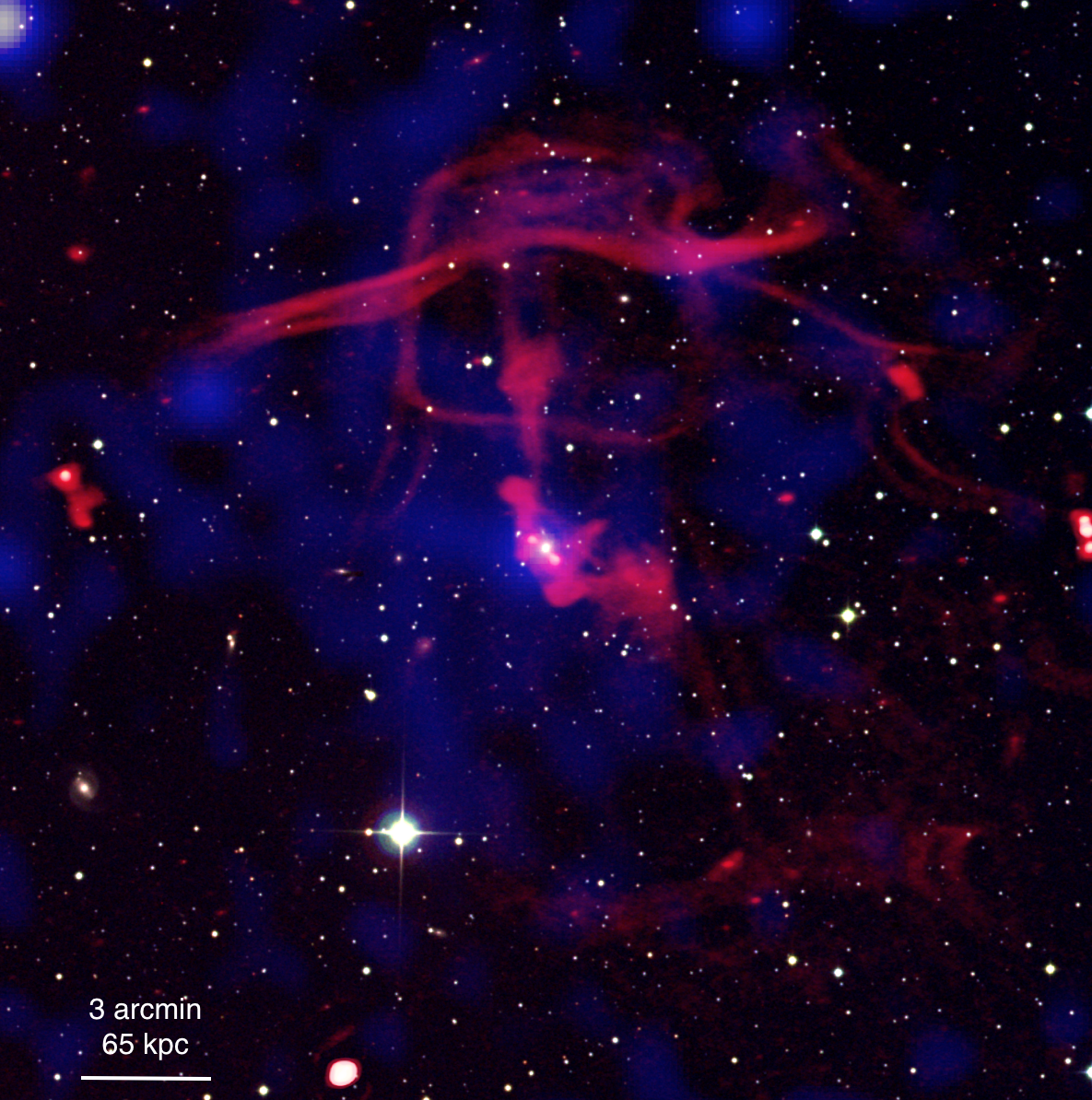}}
\captionsetup{labelformat=empty}
\caption{\textbf{Supplementary Figure 1: Composite image of the galaxy group Nest200047.} Radio data are shown in red (LOFAR image at 144 MHz with a resolution of 4.3 arcsec $\times$ 8 arcsec), X-ray data are shown in blue (SRG/eROSITA image at 0.5-2.3 keV) and optical data are shown in background (r-band, g-band and i-band Pan-STARRS images). A reference scale is shown in the bottom-left corner.}
\label{fig:rgb}
\end{figure*}

\begin{figure}[htp!]
\centering
{\includegraphics[width=11cm]{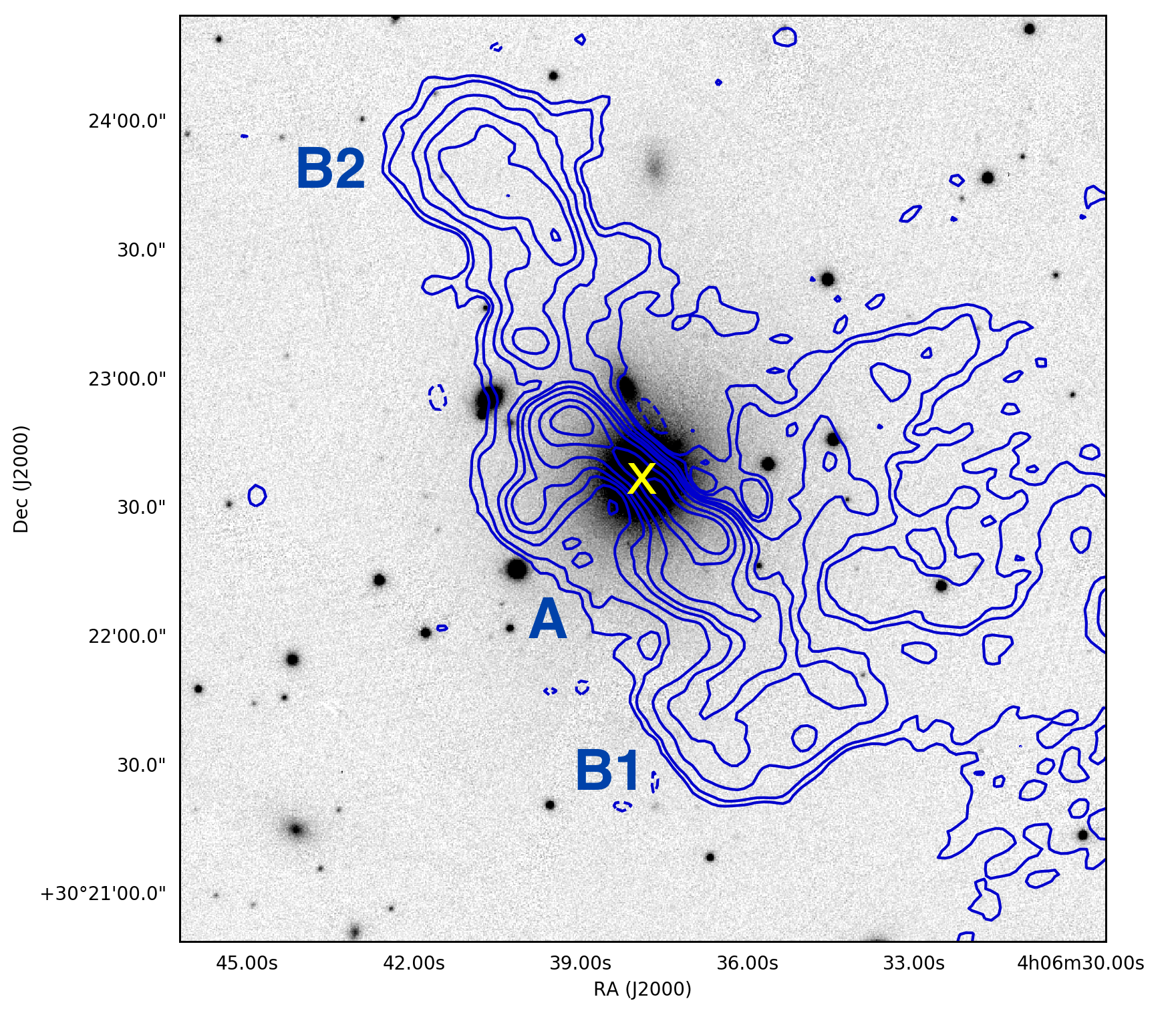}}
\captionsetup{labelformat=empty}
\caption{\textbf{Supplementary Figure 2: Zoom-in on the galaxy MCG+05-10-007 located at the centre of the galaxy group Nest200047}. LOFAR emission at 144-MHz with a resolution of 4.3 arcsec $\times$ 8.6 arcsec is shown with contours at -3, 3, 5, 10, 15, 20 $\rm \times\sigma$ ($\rm \sigma = 0.166 \ mJy \ beam^{-1}$) overlaid on the Pan-STARRS r-band image. MCG+05-10-007 is marked with a yellow cross and letters indicate the different morphological features of the radio galaxy as presented in the main text.}
\label{fig:dss}
 \end{figure}

\begin{table}[!htp]

        \small
        \captionsetup{labelformat=empty}
 \caption{\textbf{Supplementary Table 1: Details of the LOFAR observational setup at 144 MHz and 53 MHz.} Note that the target is located at $\sim$1.4 degrees from the pointing center in the 144-MHz observations, while it lies at the pointing center in the 53-MHz observations. }
        \centering
                \begin{tabular}{l c c}
                \hline
                \hline
                 & 53 MHz & 144 MHz \\
                \hline
                \hline
                Observing dates & 18-04-2020 & 28-06-2019, 03-07-2020 \\
                Bandwdith [MHz] & 30-78 MHz & 120-168\\
                Channel width [kHz] & 12.2 & 12.2\\
                Observing time [hr] & 8 & 8$\times2$\\
                Integration time [s] & 1 & 1\\
                Polarisations & 4& 4\\
                Flux calibrator & 3C196 & 3C196, 3C48\\
                 \hline
                \hline  
                \end{tabular}
     \label{tab:data}
\end{table}

\begin{table*}[!htp]

        \small
        \captionsetup{labelformat=empty}
 \caption{\textbf{Supplementary Table 2: Flux densities, luminosities and global spectral indices of different source regions.} Measurements have been performed using the LOFAR images at 53 MHz and 144 MHz with 25-arcsec resolution using the boxes shown in Figure 1, bottom-left panel.
 }
        \centering
                \begin{tabular}{c c c c c c}
                \hline
                \hline
                Region & $\rm S_{53MHz}$ & $\rm L_{53MHz}$& $\rm S_{144MHz}$ & $\rm L_{144MHz}$ & $\rm \alpha^{53MHz}_{144MHz}$\\
                & [mJy] & [$\rm \times10^{24}$ W/Hz] & [mJy] & [$\rm \times10^{24}$ W/Hz] & \\
                \hline
                \hline
                Total & 25000$\pm$2000 & 18$\pm$2 & 6800$\pm$700 & 4.8$\pm$0.5 & 1.3$\pm$0.1\\
                A+B1+B2 & 2100$\pm$200 & 1.5$\pm$0.2& 900$\pm$100 & 0.9$\pm$0.1& 0.8$\pm$0.2\\
                C1 & 1300$\pm$130 & 0.9$\pm$0.1 & 370$\pm$60 & 0.3$\pm$0.1 & 1.2$\pm$0.2\\
                C2 & 1100$\pm$100 & 0.8$\pm$0.1& 350$\pm$50 & 0.2$\pm$0.1& 1.2$\pm$0.2\\
                D1 & 2100$\pm$200 & 1.5$\pm$0.2& 520$\pm$80 &0.4$\pm$0.1& 1.4$\pm$0.2   \\
                D2 & 4800$\pm$500 & 3.4$\pm$0.4 & 1500$\pm$200 & 1.0$\pm$0.2& 1.2$\pm$0.2\\
                D3 & 2200$\pm$200 & 1.5$\pm$0.2& 560$\pm$80 & 0.4$\pm$0.1& 1.4$\pm$0.2\\
                \hline
                \hline  
                \end{tabular}
     \label{tab:flux}
\end{table*}

\begin{figure}
\centering
\includegraphics[width=14cm]{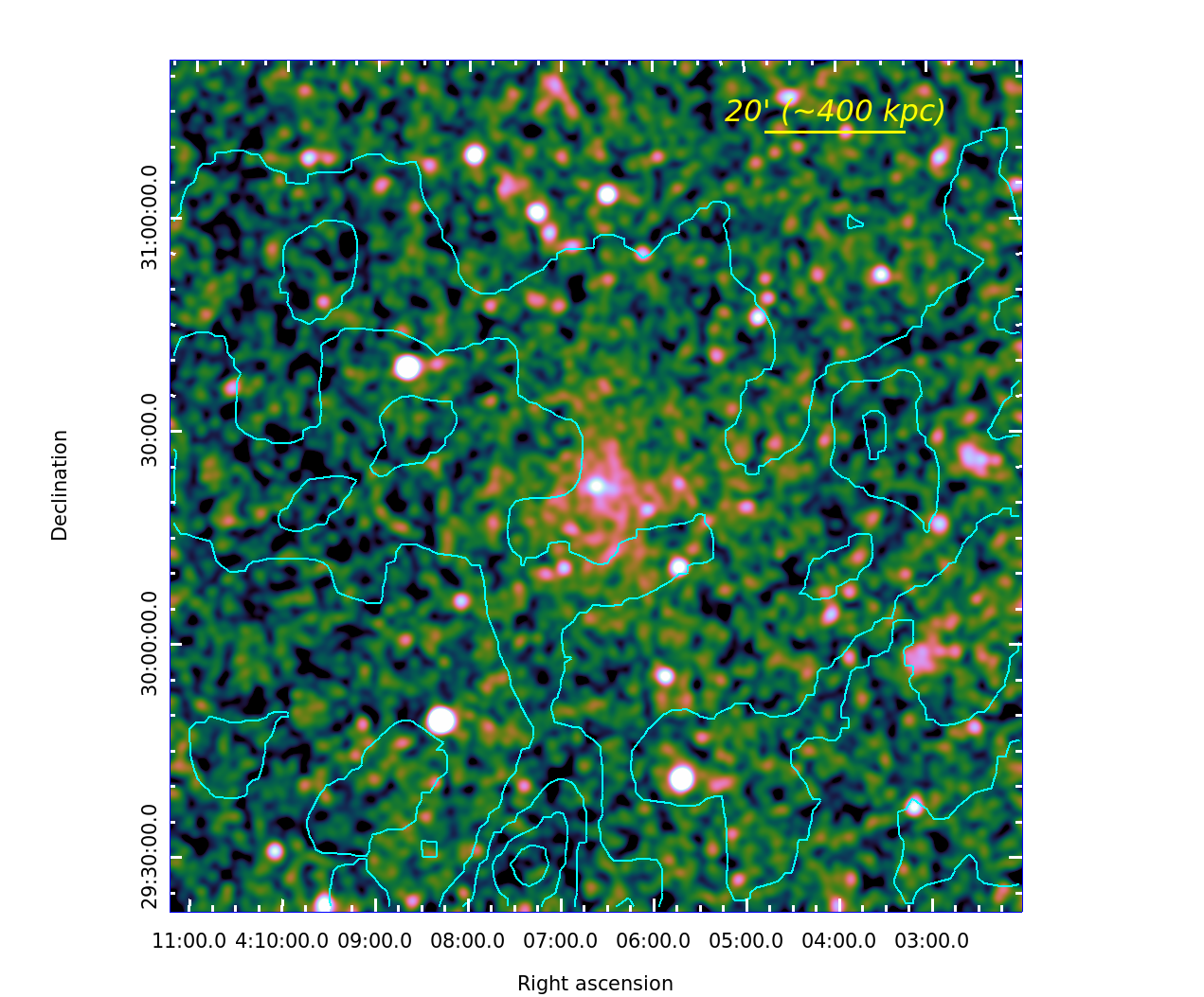}
\captionsetup{labelformat=empty}
\caption{\textbf{Supplementary Figure 3: 0.5-2.3 keV eROSITA image smoothed with a filter equal to $\sigma=40$ \mb{arcsec} to emphasize the diffuse X-ray emission of the NEST200047 group on large scales.} The emission can be traced up to $\sim 10$ arcmin from the core and is centered on the central galaxy MCG+05-10-007, which is visible as a yellow bright spot at RA 04:06:38, DEC 30:22:34. Note that due to the proximity of Nest200047 to the Galactic plane, variable low-energy photoelectric absorption across the field and variable contribution of the Galactic foreground diffuse emission might affect the appearance of this image on large scales. This is illustrated with the cyan contours, which show the values of E(B-V) from \cite{meisner2015}. The contours start at 0.25 (the lowest values in the NW corner of the image) with 0.25 increment. The highest E(B-V) value $\sim$1.6 (corresponding to NH$\sim 1.4\times10^{22}\,{\rm cm^{-2}}$) is at the Southern edge of the image. Such E(B-V) values introduce variations of the photoelectric absorption for extragalactic sources at 0.7 keV between 0.3 and $10^{-3}$ across the image shown.}
\label{fig:x_image}
\end{figure}

\begin{figure*}[!htp]
\centering
{\includegraphics[width=14cm]{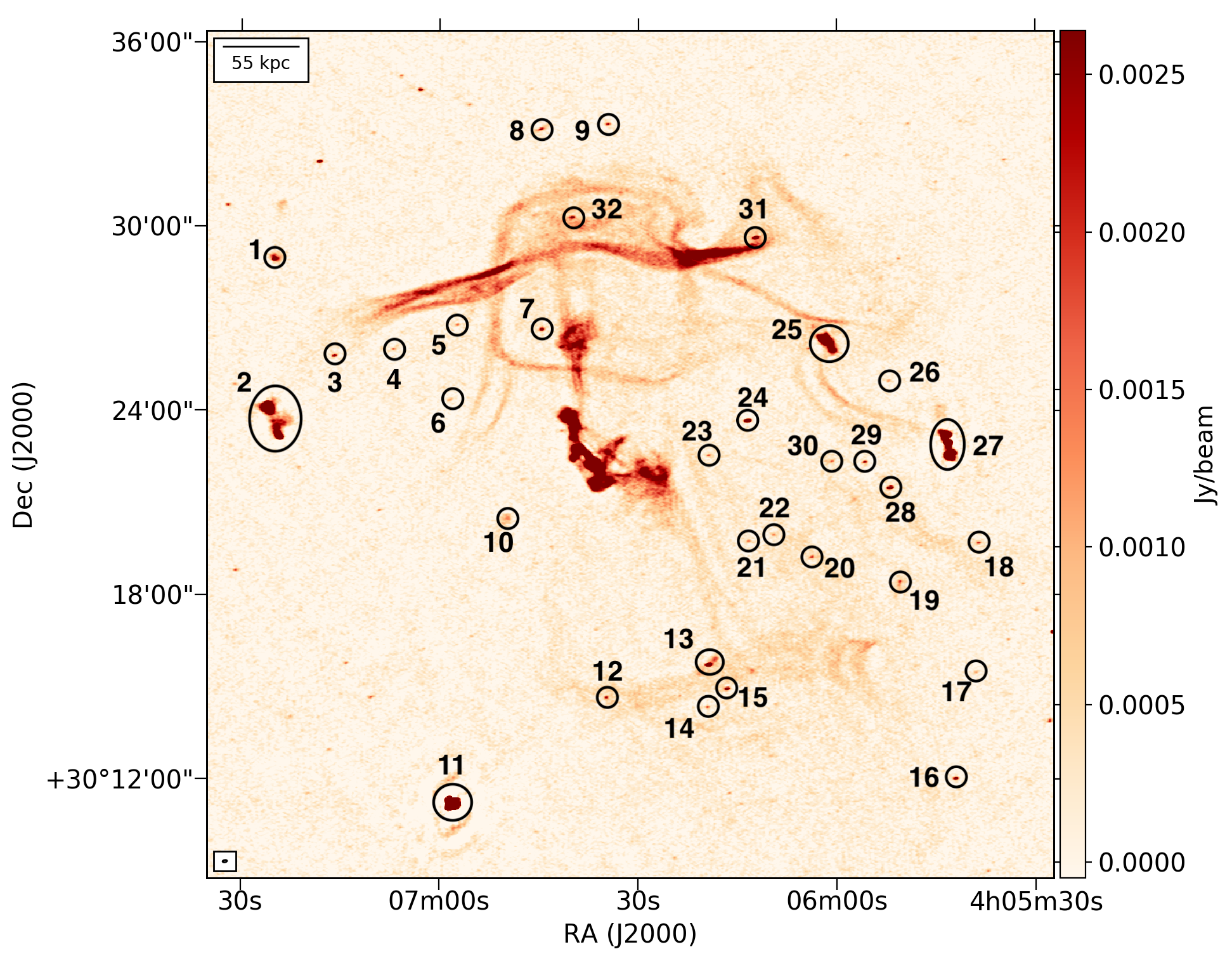}}
\captionsetup{labelformat=empty}
\caption{\textbf{Supplementary Figure 4: LOFAR image of the galaxy group Nest200047 at 144 MHz highlighting compact sources present in the field.} The LOFAR image has a resolution of 4.3 arcsec $\times$ 8.6 arcsec and a noise of $\rm \sigma = 0.166 \ mJy \ beam^{-1}$. Numbers mark all compact sources reported in Supplementary Table 3.}
\label{fig:compactsources}
 \end{figure*}

 \begin{table*}[!htp]

        \small
        \captionsetup{labelformat=empty}
 \caption{\textbf{Supplementary Table 3: Flux densities of compact sources (above 5$\sigma$) present in the field as measured from the high resolution image at 144 MHz.} All sources are marked in Supplementary Figure 4 with their respective label.}
        \centering
                \begin{tabular}{c c c c }
                
\hline
\hline
Label & RA & DEC &  $\rm S_{144MHz}$ \\
 & $\rm [J2000]$  & $\rm [J2000]$ & [mJy] \\
\hline
\hline
  1    &  04:07:24 &  +30:28:58 &17$\pm$3\\
  2    &  04:07:25 &  +30:23:45 &240$\pm$40\\
  3    &  04:07:15 &  +30:25:46 &3.2$\pm$0.6\\ 
  4    &  04:07:06 &  +30:25:59 &1.6$\pm$0.5\\ 
  5    &  04:06:57 &  +30:26:47 &1.6$\pm$0.5 \\
  6    &  04:06:58 &  +30:24:20 &1.2$\pm$0.5 \\
  7    &  04:06:44 &  +30:26:38 &5$\pm$1\\
  8    &  04:06:44 &  +30:33:07 &4$\pm$1\\	 
  9    &  04:06:34 &  +30:33:20 &3$\pm$1 \\
  10   &  04:06:50 &  +30:20:28 &3$\pm$1 \\
  11   &  04:06:58 &  +30:11:12 &120$\pm$20\\
  12   &  04:06:34 &  +30:14:39 &3$\pm$1 \\
  13   &  04:06:19 &  +30:15:43 &30$\pm$5 \\
  14   &  04:06:19 &  +30:14:20 &1.6$\pm$0.5 \\
  15   &  04:06:16 &  +30:14:56 &6$\pm$1 \\
  16   &  04:05:42 &  +30:12:00 &4$\pm$1\\
  17   &  04:05:39 &  +30:15:27 &1.2$\pm$0.5\\ 
  18   &  04:05:38 &  +30:19:41 &2.2$\pm$0.5 \\
  19   &  04:05:50 &  +30:18:19 &3$\pm$1\\
  20   &  04:06:03 &  +30:19:13 &2.2$\pm$0.5\\ 
  21   &  04:06:13 &  +30:19:45 &1.5$\pm$0.5 \\
  22   &  04:06:09 &  +30:19:57 &1.5$\pm$0.5 \\
  23   &  04:06:19 &  +30:22:33 &1.9$\pm$0.5 \\
  24   &  04:06:13 &  +30:23:40 &17$\pm$3 \\
  25   &  04:06:01 &  +30:26:17 &110$\pm$20\\
  26   &  04:05:52 &  +30:24:56 &1.4$\pm$0.5 \\
  27   &  04:05:43 &  +30:22:57 &400$\pm$60\\
  28   &  04:05:51 &  +30:21:28 &10$\pm$2 \\
  29   &  04:05:55 &  +30:22:18 &2.8$\pm$0.6 \\
  30   &  04:06:00 &  +30:22:21 &1.7$\pm$0.5 \\
  31   &  04:06:12 &  +30:29:38 &7$\pm$1 \\
  32   &  04:06:39 &  +30:30:12 &5$\pm$1 \\

                \hline
                \hline  
                \end{tabular}
     \label{tab:unrelated-sources}
\end{table*}

\begin{figure}
\centering
{\includegraphics[width=14cm]{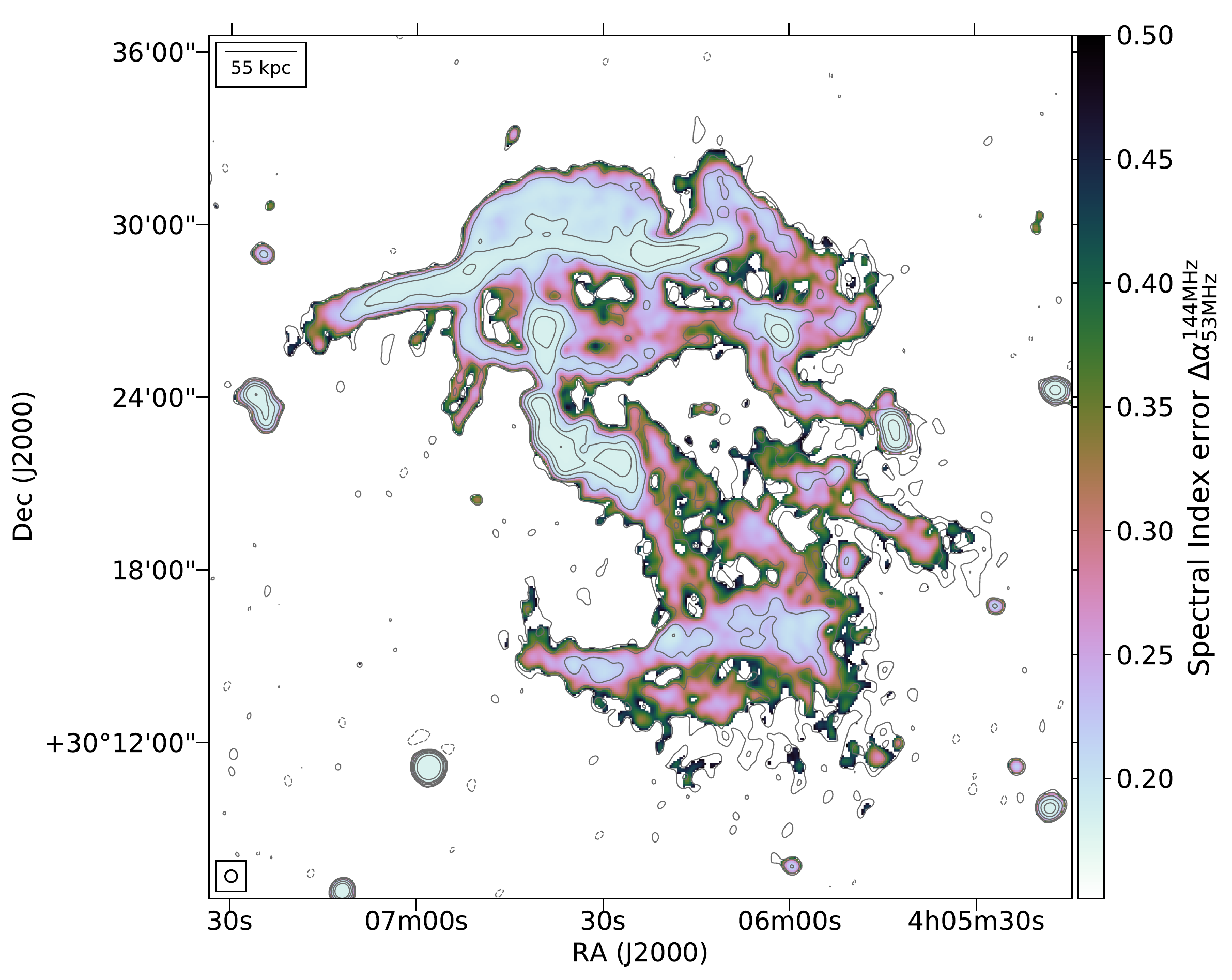}}
\captionsetup{labelformat=empty}
\caption{\textbf{Supplementary Figure 5: Spectral index $1 \sigma$ error map of the galaxy group Nest200047 obtained using the LOFAR images at 53 MHz and 144 MHz at 25 arcsec resolution.} Contour levels represent the emission at 53 MHz and are drawn at -3, 3, 5, 10, 20, 35, 100 $\times \sigma$ levels, with $\sigma$=2.7 mJy $\rm beam^{-1}$. The beam size is shown in the bottom left corner of the image.}
\label{fig:spix_error}
 \end{figure}

\begin{figure}
\centering
\includegraphics[width=11cm,trim= 0cm 3cm 0cm 3cm]{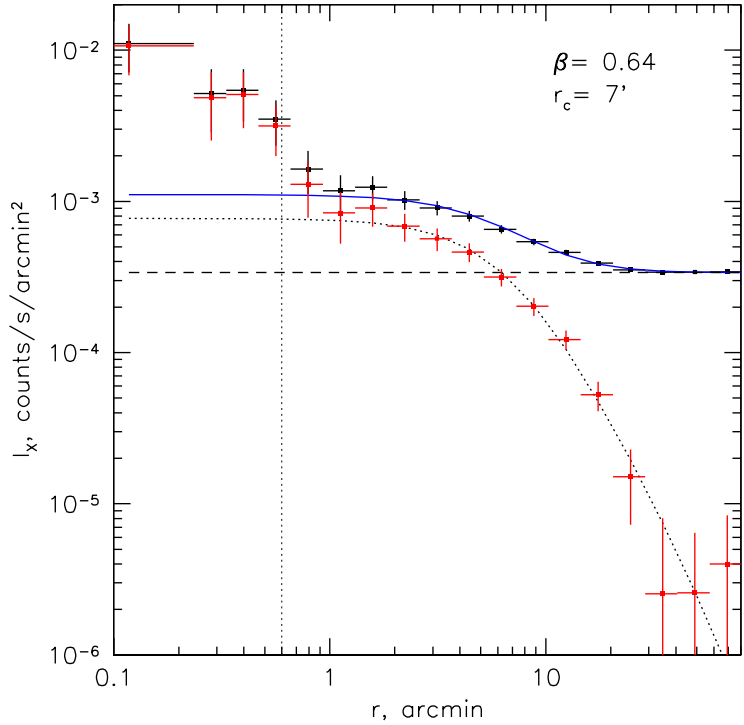}
\captionsetup{labelformat=empty}
\caption{\textbf{Supplementary Figure 6: Surface brightness profile of the galaxy group Nest200047 in the 0.5-2.3 keV band (black
points).} The blue curve shows the best-fitting $\beta$-model plus a constant background (depicted with horizontal dashed line).  
The red points show the same data from which the contribution of the background has been subtracted. For the black and red points, the vertical uncertainties correspond to $1\,\sigma$ statistical errors. For comparison, the dotted black line shows the pure $\beta$-model. The bar-like asymmetric X-ray structure near the center (see Figure 2) has been excluded from the fit (left from the dotted vertical line).}
\label{fig:x_profile}
\end{figure}

 \begin{table}[!htp]

        \small
        \captionsetup{labelformat=empty}
 \caption{\textbf{Supplementary Table 4: Physical properties of the main structures of the source.} Column 1: region name; columns 2-3; ellipsoidal semiminor and semimajor axis used to compute the volume; column 4: volume; column 5-6: magnetic field and non-thermal pressure computed assuming minimum energy condition.}
        \centering
                \begin{tabular}{c c c c c c}
                \hline
                \hline
              Region & a & b & Volume & $\rm B_{min}$ & $\rm p_{nth}$ \\
               & [kpc] & [kpc] & [cm$^3$] & [$\mu G$] & [$\rm dyne \ cm^{-2}]$\\
                 \hline
                \hline
                C1 & 16 & 31 & 1$\rm \times 10^{69}$ &4.2& $4.3 \times 10^{-13}$\\
                C2 & 16 & 31 & 1$\rm \times 10^{69}$ & 4.2 & $4.2 \times 10^{-13}$\\
                D1 & 24 & 100 & 7.5$\rm \times 10^{69}$&  3.0 & $2.1 \times 10^{-13}$\\
                D2 & 17 & 158 & 6$\rm \times 10^{69}$& 3.9 & $3.7 \times 10^{-13}$\\
                D3 & 35 & 75 & 1$\rm \times 10^{70}$ & 2.8 & $1.8 \times 10^{-13}$\\
                \hline
                \hline  
                \end{tabular}
     \label{tab:equipart}
\end{table}

\begin{table}[!htp]

\small
\captionsetup{labelformat=empty}
\caption{\textbf{Supplementary Table
5: Bubble properties.} Column 1: bubble name; column 2: projected distance of the bubble from the group center; column 3: bubble volume assuming ellipsoidal geometry; column 4: sound-speed time; column 5: buoyancy time; colum 6: bubble power computed as $\rm p_{th}V/t_{buoy}$, with $\rm p_{th}=1-4\times 10^{-12} \ dyne \ cm^{-2}$.}

\centering
                \begin{tabular}{l c c c c c}
                \hline
                \hline
                 Bubble & $H$ & $V$ & $t_{cs}$ & $t_{buoy}$ & $P_{bubble}$ \\
                 &  [kpc] & [$\rm cm^3$] & [Myr] & [Myr] & [erg/s]\\ 
                \hline
                \hline
                C2 &  90 & 1$\rm \times10^{69}$ & 120 & 130 & 3$\rm \times10^{41}$ - 1$\rm \times10^{42}$\\
                D3 & 170 & 1$\rm \times10^{70}$ & 240 & 350 & 1-4$\rm \times10^{42}$\\
                 \hline
                \hline  
                \end{tabular}
     \label{tab:bubbles}
\end{table}

\bibliography{nest200047.bib}

\end{document}